\documentclass[preprint,showpacs,prl,superscriptaddress,aps]{revtex4-1}
\usepackage{amssymb}
\usepackage{graphicx}
\usepackage{color}
\usepackage{epstopdf}
\usepackage{ulem}

\begin{document}

\author{Cong Mai}
\affiliation{Department of Physics, North Carolina State University, Raleigh, NC 27695, USA}
\title{Temperature Dependent Valley Relaxation Dynamics in Single Layer WS$_{2}$ Measured Using Ultrafast Spectroscopy}
\author{Yuriy G. Semenov}
\affiliation{Department of Electrical and Computer Engineering, North Carolina State University, Raleigh, NC 27695, USA}
\author{Andrew Barrette}
\affiliation{Department of Physics, North Carolina State University, Raleigh, NC 27695, USA}
\author{Yifei Yu}
\affiliation{Department of Material Science and Engineering, North Carolina State University, Raleigh, NC 27695, USA}
\author{Zhenghe Jin}
\affiliation{Department of Electrical and Computer Engineering, North Carolina State University, Raleigh, NC 27695, USA}
\author{Linyou Cao}
\email{lcao2@ncsu.edu}
\affiliation{Department of Material Science and Engineering, North Carolina State University, Raleigh, NC 27695, USA}
\author{Ki Wook Kim}
\email{kwk@ncsu.edu}
\affiliation{Department of Electrical and Computer Engineering, North Carolina State University, Raleigh, NC 27695, USA}
\author{Kenan Gundogdu}
\email{kgundog@ncsu.edu}
\affiliation{Department of Physics, North Carolina State University, Raleigh, NC 27695, USA}

\begin{abstract}

We measured the lifetime of optically created valley polarization in single layer WS$_{2}$ using transient absorption spectroscopy. The electron valley relaxation is very short ($<1$ps). However the hole valley lifetime is at least two orders of magnitude longer and exhibits a temperature dependence that cannot be explained by single carrier spin/valley relaxation mechanisms. Our theoretical analysis suggests that a collective contribution of two potential processes may explain the valley relaxation in single layer WS$_{2}$. One process involves direct scattering of excitons from $K$ to $K^\prime$ valleys with a spin flip-flop interaction. The other mechanism involves scattering through spin degenerate $\Gamma$ valley. This second process is thermally activated with an Arrhenius behavior due to the energy barrier between $\Gamma$ and $K$ valleys.

\end{abstract}

\maketitle

\section{\protect\bigskip Introduction}

The discovery of graphene, a monolayer of carbon atoms, has inspired considerable interest in other 2D material systems in search of exotic electronic, optical and mechanical properties and novel practical applications\cite{Novoselov05,Geim09,Butler13,JariwalaDevices}. In particular, two-dimensional transition metal dichalcogenides (TMDC) recently emerged as a promising material system with many potential electronic applications. With their tunable direct gap in visible range of the optical spectrum, absence of dangling bonds and high surface-to-volume ratio, these 2D semiconducting systems are ideal for field-effect transistors (FET), photovoltaics, light emitting diodes (LEDs), molecule sensing, and electrocatalytic water splitting applications\cite{Wang12,Radisavljevic11,Britnell13,Sundaram13,Perkins13,He13,Splendiani10,Mak10,Xiao12,YuEWS}.
Moreover, due to strong spin-orbit splitting they have been subject to specific spin and valley applications. The optical band gap in these structures are located at the $K$ and $K^{\prime }=-K$ points at the edge of the Brillouin zone. A strong spin-orbital coupling results in distinct states with different spin and valley indices so that $K$ to $K^{\prime }$ elastic transition requires a spin flip for the electrons and holes. One immediate consequence of this property is the ability to control valley polarization, hence crystal quasi-momentum of electrons, by using circularly polarized light, which is impossible in conventional semiconductors due to negligible momentum of photons. This could open up opportunities for developing optoelectronic and valleytronic applications based on manipulation of spin and valley polarization of charge carriers\cite{Splendiani10,Mak10,Xiao12,Mak12,Zeng12,Cao12}.

Detailed understanding of inter-valley relaxation dynamics is critical for the implementation of valleytronic applications. Very recently, numerous experimental and theoretical efforts have been devoted to characterization of electronic structure and optical properties of TMDCs\cite{Kosmider13,Gutierrez13,Zeng13,Zhu11,Zhao13,Jones13,Sun13,Shi13,Sim13}.  However, there are only limited experimental studies that directly address the valley and spin relaxation process.  Recently we measured valley lifetime in single layer MoS$_{2}$ to be 10 ps at 74K, using time-resolved absorption spectroscopy\cite{Mai14}. Surprisingly this relaxation time is very short, as large spin splitting in the valence band and spin valley coupling in $K$ and $K^{\prime }$  valleys was expected to impede hole valley scattering\cite{Splendiani10,Mak10,Xiao12,Mak12,Zeng12,Cao12,Sim13,Mai14}. As of now an accurate picture of valley relaxation mechanisms in atomically thin TDMCs is missing. In the current work, we studied thermal dependence of valley relaxation of excitons in monolayer WS$_{2}$  using broadband transient absorption spectroscopy. In WS$_{2}$  the spin-orbit splitting is significantly larger compared to MoS$_{2}$ and other TMDCs \cite{Xiao12,Kosmider13,Gutierrez13,Zeng13,Zhu11}.
Even so we observe that hole valley lifetime is more than an order of magnitude larger in WS$_{2}$ compared to MoS$_{2}$. Similar to MoS$_{2}$, electron valley relaxation is dominated by many body interactions and faster compared to hole valley relaxation. Based on temperature dependence of valley lifetimes, we propose a mechanism for an exciton-mediated valley relaxation process for holes in single layer WS$_{2}$.

In order to measure the valley relaxation time, we performed experiments on single layer WS$_{2}$ samples grown on quartz substrates using a chemical vapor deposition technique (Sup. Inf.)\cite{YuVapordep,S1,S2,S3}. In time resolved experiments, circularly polarized 60 fs pump pulses, tuned to lower energy tail of A excitonic transition (1.977 eV) to create electron-hole pairs in the K valley. The differential transmission of a same circularly polarized (SCP) broad band white-light continuum probe pulse measures the relaxation within the same valley and opposite circularly polarized probe pulse (OCP) measures the population in the other valley. The decay of the polarization anisotropy between the SCP and the OCP spectra reveals the intervalley relaxation dynamics. 

Figure 1 (a) schematically displays the electronic band structure for WS$_{2}$ at the $K$ and $K^{\prime }$ points, adapted from Ref. \cite{Sallen12}. The conduction band is composed by $d[1-3z^{2}]$ orbitals $\left\vert L,m\right\rangle$ of $W$ with zero magnetic quantum number $m=0$ under orbital moment $L=2$ and relatively small admixture of $p[x-i\nu y]$ orbitals of $S$ \cite{Cao12} with $m=-\nu$, which constitutes relatively small, 27 meV, spin splitting ($\nu =\pm 1$ corresponds to $K$ and $K^{\prime }$ valleys). The valence band is formed by $d[(x+i\nu y)^{2}]$ orbitals of $W$  with $m=2\nu $ and some admixture of $p$-orbitals of $S$ that conditions much larger, 435 meV, valence band spin splitting \cite{Cao12,Kosmider13}. This spin splitting leads to energetically well separated $A$ and $B$ excitonic transitions shown by red and blue arrows in Fig. 1a. Their electro-dipole interband optical transitions are coupled with circularly polarized light. 

Figure 1(b) shows transient SCP and OCP spectra for $A$ exciton at 110 K. Initially the SCP spectrum shows a dispersive line shape in which the lower energy tail of the spectrum exhibits ground state bleaching (GSB) and stimulated emission (SE) and the higher energy tail exhibits photoinduced absorption (PIA). This dispersive feature evolves into purely absorptive line shape in about 7.6 ps. The OCP spectrum also exhibits a dispersive line shape, though less prominent than that of SCP. In the later time delays the anisotropy between the SCP and OCP spectra vanishes and a sharp dip shows up at 2.04 eV.

We attribute this dispersive distortion to phase-space filling effects similar to those observed in conventional semiconductors\cite{Lee86,Christodoulides10}. Briefly the photoexcited population blue-shifts the A exciton absorption, resulting in a differential line-shape in the transient spectrum. Monolayer materials exhibit ultra-tight quantum confinement\cite{Zeng13,Zhu11,Christodoulides10}.  Thus phase-space filling of carriers, which is related to Pauli-blocking, is very significant\cite{Sim13,Lee86,Christodoulides10}. A similar blue shift has been observed in MoS$_{2}$ transient absorption studies as well\cite{Sim13}.  The exact source of the spectral dip at 2.04 eV is unclear. It could be due to an overlapping absorptive transition from the excited state. A better understanding of the band structure and the excited states of the WS$_2$ is needed for analyzing this feature. 

The evolution of the polarization anisotropy, measured by the difference between the SCP and OCP spectra in Figure 1, reveals the valley relaxation time of electrons and holes. The immediate presence of $A$ exciton feature in the OCP spectra clearly suggests that a significant fraction of electron population, initially photo-excited in the $K$ valley, quickly delocalize between $K$ and $K^{\prime }$ valleys within 200 fs, which is within the time resolution of the experiment. This is because without any valley relaxation OCP should not exhibit any $A$ exciton feature. The electron relaxation into the $K^{\prime }$ valley (i.e. transition of the direct exciton [$eK,hK$], with both carriers in K-valley, to indirect one (dark exciton) [$eK^{\prime },hK$]) leads to bleaching of the $A$ exciton absorption in the OCP spectra. This immediate electron valley delocalization has been observed in single layer MoS$_{2}$ and attributed to coherent coupling of excitonic states\cite{Sim13}. While the circularly polarized optical transitions are valley specific, the resulting optically created exciton associated with a specific valley, $K$ or $K^{\prime }$ , is not an eigenstate of the full exciton problem. Strong electron-hole confinement in exciton with energy 710 meV \cite{Peyghambarian84} and radius around 1 - 2 nm couples the $K$ and $K^{\prime }$ valleys. As a result the electron in $K$ valley quickly delocalizes over both valleys. This is further confirmed with the analysis of the evolution of the $B$ exciton transition. As observed in Figure 2 (a, b), the $B$ transition bleached immediately after the pump pulse in both SCP and OCP spectra with equal intensity. Because the $B$ valence levels are very high in energy and are not populated with the pump pulse, it is only sensitive to electron dynamics. Hence electrons occupy $B$-levels in both valleys immediately upon photoexcitation. Because $B$-levels have opposite spin orientation; these spectra suggest not only that valley relaxation of the electrons is very quick but also that electron spin relaxation within the same valley is very fast (Fig. 2 c, d).

In contrast to fast conduction band spin and valley relaxation, hole delocalization is not favorable energetically due to the large, $\Delta
_{SO}\simeq $450 meV, spin orbital splitting, which imposes the energy gap between the parallel spin states in the two valleys (Fig. 1a). Hence hole valley relaxation takes a much longer time. The total loss of the polarization anisotropy in the SCP and OCP spectra show that it is around 100 ps. We note that this is much longer compared to MoS$_{2}$ where the valley relaxation takes place in about 10 ps. 

In order to resolve hole valley relaxation mechanism we performed transient absorption experiments in a range of temperatures from 74 K to 298 K. For each temperature, the pump pulses were tuned to the lower energy side of the $A$ excitonic transition and resulting dynamics were probed with a broadband white light continuum pulse. At each temperature the early spectra show similar dynamics to 110 K data suggesting fast spin and valley relaxation dynamics for the electrons. However relaxation of the circularly polarized anisotropy clearly evolves at a different rate, revealing the thermal dependence of the hole valley relaxation dynamics. 

In order to quantitatively extract the hole valley relaxation time, we analyzed the decay of polarization anisotropy at the $A$ exciton transition by taking the difference of SCP and OCP spectra\cite{Zhu14}. Neither single nor double exponential decay functions fit the anisotropy decay. Therefore we used a triple exponential fit. For all temperatures the first time constant is in 100 fs range, within the resolution of the experiment. The second time constant is in 1.5-3 ps range (Sup. Inf.). The slow component is the only one that exhibits clear temperature dependence varying from 88 ps to 8 ps at 74 K and 298 K respectively (Fig. 3). In the following this slow component is used for theoretical analysis of the valley relaxation process.

We considered several previously proposed inter-valley relaxation mechanisms to explain the experimental results. Our initial analysis suggests single carrier relaxation mechanisms cannot be responsible for the temperature dependence that we observe. For instance scattering with nonmagnetic impurities has been suggested as a potential valley relaxation process. Reference \cite{Lu13} shows that this mechanism requires a $\sim k^{4}$ dependence on hole momentum, which leads to a stronger temperature dependence ($T^2$) than our observation. Relaxation through flexural phonon modes was also considered as a potential mechanism\cite{Song13}. It predicts an inverse relation between mobility and relaxation rate, and order of magnitude longer valley lifetimes compared to our results, hence it is unlikely to be responsible for the valley relaxation in WS$_{2}$. 

We considered spin relaxation mechanisms such as Elliot-Yafet (EY) and Dyakonov-Perel (DP) processes as well. These processes require scattering of carriers under inhomogeneous magnetic field. For instance in conventional semiconductors, Dressalhause effect leads to such an effective field. Unlike many conventional semiconducting materials in TDMCs the spin quantization axis retains normal direction over all of Brillouin zone except the specific points where spin splitting reduces to zero. Such a property is clearly demonstrated in Ref. \cite{Cao12} based on calculations involving 80 energy bands of MoS$_{2}$. Our first principle calculations for WS$_{2}$ also reproduce the absence of transversal spin components over all of Brillouin zone. Thus the hole/electron diffusive motion cannot mix the states with opposite spin directions along the normal to layer plane (i.e. z axis).  Therefore Elliot-Yafet and Dyakonov-Perel mechanisms are irrelevant.

Note however that in a realistic structure the spin quantization axis deviates from the normal direction due to inversion asymmetry induced by the substrate. Indeed, modeling the substrate induced asymmetry by an effective external electric field, we found the spin quantization axis deviates from z-direction. Proportionality of this deviation to quasi-momentum shift from extremum points represents the Rashba effect with Hamiltonian $H_{R}=\alpha _{\varphi }\left\vert \Delta
\mathbf{k}\right\vert $, ($\Delta \mathbf{k=\pm K-k}$, $\varphi $ is an
angle between $\Delta \mathbf{k}$ and in-plane axis $x$), which in addition reveals trigonal anisotropy of $\alpha _{\varphi }$ in the vicinities of each $\mathbf{K}$ and $\mathbf{K}^{\prime }=-\mathbf{K}$ points. The correspondent spin-relaxation rate is,%
\begin{equation}
\tau _{s}^{-1}=\frac{2}{\hbar ^{2}}\left\langle \frac{\alpha _{\varphi
}^{2}\Delta \mathbf{k}^{2}\tau _{k}}{\tau _{k}^{2}\omega _{SO}^{2}+1}\right\rangle  \label{t}
\end{equation}%

where brackets mean thermal averaging, $\omega _{SO}=\Delta _{SO}/\hbar $,
and $\tau _{k}$ is  momentum relaxation time. Our first principle calculations estimate the Rashba constant mediated by potential drop $\Delta U$  between two sides of a layer. Averaging over angle $\varphi$, we find $\alpha =\beta \Delta U$, where $\beta \simeq 1.1\cdot 10^{-10}$ cm, which results in an insignificant contribution ($\tau _{s}^{-1}\sim 10^{-16}\left\langle \tau _{k}^{-1}\right\rangle $) of this mechanism to intervalley spin relaxation even at $\Delta U$=100 meV.

 We conclude that the previously discussed single particle relaxation mechanisms are not sufficient to explain exciton intervalley relaxation. On the other hand, the excitonic electron-hole exchange interaction lifts orthogonality of spin states attributed to different valleys. This effect describes a minimal exciton spin Hamiltonian

\begin{equation}
H_{0}=\frac{\tau }{2}\Delta _{SO}\sigma _{z}+\frac{\tau }{2}\delta
_{SO}s_{z}+\Delta _{eh}\mathbf{\sigma s,}  \label{1}
\end{equation}%
where $\mathbf{\sigma }$ and $\mathbf{s}$ are the Pauli matrixes for electron and hole spins, $\tau =\pm 1$ the valley index, $\Delta _{SO}$ and $\delta _{SO}$ are the hole and electron spin splitting induced by spin-orbital interaction and $\Delta _{eh}$ is the strength of electron-hole exchange interaction. It is important to note that this interaction mixes the exciton spin states located in different valleys. This is because electro-dipole interband transitions occur between electronic states with same spin that generates electron-hole pair with opposite spin directions, i.e. with zero projections $s_{z}+\sigma _{z}=0$. Thus the transitions between $K$ and $-K$ excitons would be proportional to the spin flip-flop factor $T=\left\vert \left\langle \uparrow ,\downarrow \left\vert\downarrow ,\uparrow \right. \right\rangle \right\vert ^{2}$, where $\left\vert \downarrow ,\uparrow \right\rangle $ corresponds to $s_{z}=-1/2$. If the spin-independent intervalley relaxation rate is $\tau _{KK^{\prime }}^{-1}$ the net exciton relaxation rate will be

\begin{equation}
\tau _{ex}^{-1}=\tau _{KK^{\prime }}^{-1}\frac{4\Delta _{eh}^{2}}{(\Delta
_{SO}-\delta _{SO})^{2}+4\Delta _{eh}^{2}}.  \label{2}
\end{equation}

Exciton scattering on any (not only spin-associated) local defects or phonons actualize such transitions. In contrast to free carriers, excitons are composed of band states involving those far away from the extramums $K$ or $K^{\prime }$, due to their short radius. These states admix the conduction band state $\left\vert 2,0\right\rangle$ to valence bands of both valley states which lifts their orthogonality. For instance, small deviation∼ $\sim0.1\left\vert K-K^{\prime }\right\vert$ from extremum leads to non-negligible admixture of conduction band by a factor of about $\xi \sim 0.01$.

In addition to the excitonic flip-flop mechanism, carrier scattering through $\Gamma $-point could also contribute to the valley relaxation. Here the optically excited $K-$exciton scatters to an indirect one $[K_{e}(K_{e}^{\prime })\Gamma _{h}]$ and then to direct $K^{\prime }-$exciton. This mechanism is not subjected to spin restrictions with factor in Eq. (\ref{2}) due to zero spin-orbital splitting ($\Delta _{SO}=\delta _{SO}=0$) at the $\Gamma $ point. Moreover, since $\Gamma $ state is composed from $\left\vert 2,0\right\rangle $ and $\left\vert 1,0\right\rangle $ orbitals, the $K-\Gamma $ overlap is substantially stronger than direct inter-valley overlap. On the other hand the difference $\Delta _{K\Gamma}^{ex}$ in direct exciton energy $E_{ex}[K_{e}K_{h}]=E_{e}(K)-E_{v}(K)-E_{X}(K)$ and indirect one $E_{ex}[K_{e}\Gamma _{h}]=E_{e}(K)-E_{v}(\Gamma )-E_{X}(K_{e}\Gamma _{h})$
imposes a thermal activation process with Arrhenius-like temperature
dependence

\begin{equation}
\tau _{K\Gamma }^{-1}=r_{K\Gamma }\exp (-\Delta _{K\Gamma }^{ex}/kT),
\label{3}
\end{equation}

where $\Delta _{K\Gamma }^{ex}=E_{v}(K)-E_{v}(\Gamma)+E_{X}(K)-E_{X}(K_{e}\Gamma _{h})$, the $E_{e(v)}(B)$ is the energy at the
conduction (valence) $B-$band edge $[B=K,\Gamma ]$ and $E_{X}(K)$ and $E_{X}(K_{e}\Gamma _{h})$ are the binding energies for direct and indierct
excitons. Our first principal calculations (Sup. Inf.) give $E_{v}(K)-E_{v}(\Gamma )=310$meV, which is slighly differerent from previous calculations\cite{Zhu11}. In the 2D Wannier-Mott exciton model the difference  $E_{X}(K)-E_{X}(K_{e}\Gamma _{h})=\Delta E_{X}$ basically stems from different reduced effective masses $\mu (K)$ and $\mu (\Gamma )$ so that $\Delta E_{X}=E_{X}(K)[1-\mu
(\Gamma )/\mu (K)]$\cite{Zhu14}.

We estimated the reduced effective masses $\mu (K)$ and $\mu (\Gamma )$ based on our first principle energy band structure calculations (Sup. Inf.)\cite{S4,S5,S6}. The resulting binding energy difference between the direct $K-$exciton and indirect  $[K_{e}(K_{e}^{\prime })\Gamma _{h}]$ exciton, $\Delta E_{X}\simeq -170$ meV, gives an energy barrier of $\Delta _{K\Gamma }^{ex}\simeq 140$ meV.

Quantitative estimation of $r_{K\Gamma }$ , $\Delta _{eh}$ and $\tau
_{KK^{\prime }}^{-1}$ would be grounded on theory of exciton formation in
TMDCs, which is not completed yet. Therefore we will treat these parameters
as phenomenological ones. We observe that low-temperature thermal dependence of $\tau
_{KK^{\prime }}^{-1}$ (Fig. 3b) is similar to that of
phonon-assistant momentum relaxation rate $\tau _{p}^{-1}=\tau _{p}^{-1}(T)$ calculated from first principles (Sup. Inf.)\cite{S7}. For free holes $\tau _{h}^{-1}$ can be
approximated with $\tau _{h}^{-1}(T)=r_{h}[1+(T/T_{0})]$ at $r_{h}=7.5\cdot $
ps$^{-1}$ and $T_{0}\simeq 200$ K in wide range (70 K to 200 K) of
temperatures. Adapting similar dependence for exciton momentum relaxation
and combining $\tau _{e,h}^{-1}(T)$ with Eq. (\ref{3}) the net result of
both mechanisms describes the temperature dependence in the form
\begin{equation}
\tau _{ex}^{-1}=r_{KK^{\prime }}(1+T/T_{0})+r_{K\Gamma }\exp (-\Delta
_{K\Gamma }^{ex}/kT).  \label{4}
\end{equation}
Fig. 3b shows that only two free parameters, $r_{KK^{\prime }}=0.01$ ps$^{-1}$
and $r_{K\Gamma }=24$ ps$^{-1}$, describe all data with experimental
accuracy.

To conclude, we present ultrafast valley relaxation dynamics measurements in monolayer WS$_{2}$. The intervalley scattering lifetime in monolayer WS$_2$ is much longer than that of monolayer MoS$_{2}$. The thermal dependence of the valley relaxation rate indicates that, due to strong exciton binding energy and exchange interaction, electron-hole spin flip-flop mechanism becomes an efficient spin relaxation channel. Because such excitonic relaxation mechanisms will not affect free carrier valley relaxation, we predict much longer valley lifetime for free carriers, which is important for future efforts in developing spintronic/valleytronic devices based on TMDCs.

 KWK acknowledges the support from SRC/NRI SWAN.

\clearpage

\begin{center}
\begin{figure}[tbp]
\includegraphics[width=14cm,angle=0]{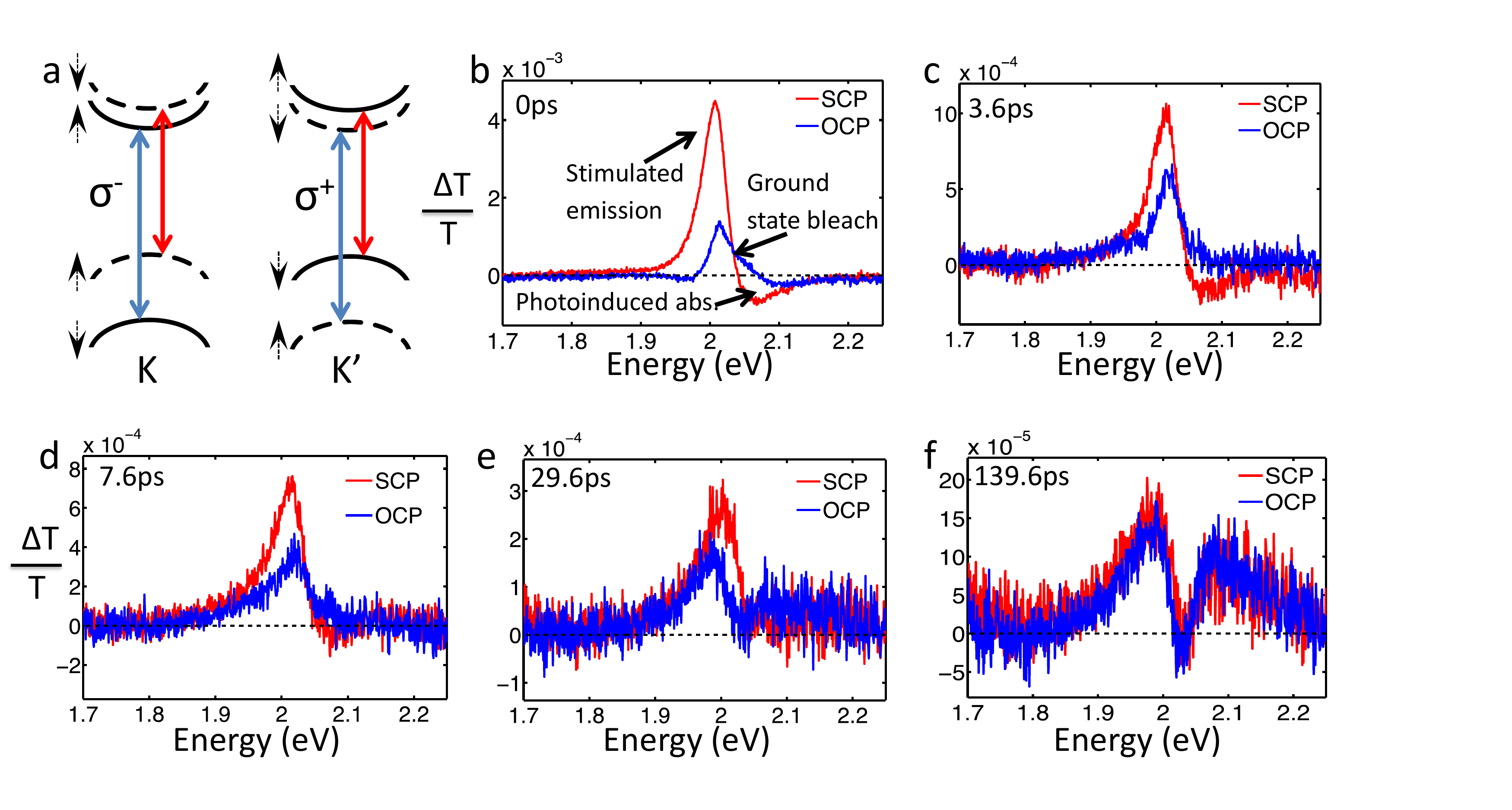}
\caption{(Color online) (a) Electronic structure and the optical transition at the $K$ and $K^\prime$ points of monolayer WS$_2$. The
dashed(solid) curves correspond to different spin states in A and B excitonic transitions. (b-f) Differential transmission at various delay at 110K for same (red) and opposite (blue) circularly poalrized pump and probe pulses.}
\end{figure}
\end{center}

\clearpage

\begin{center}
\begin{figure}[tbp]
\includegraphics[width=12cm,angle=0]{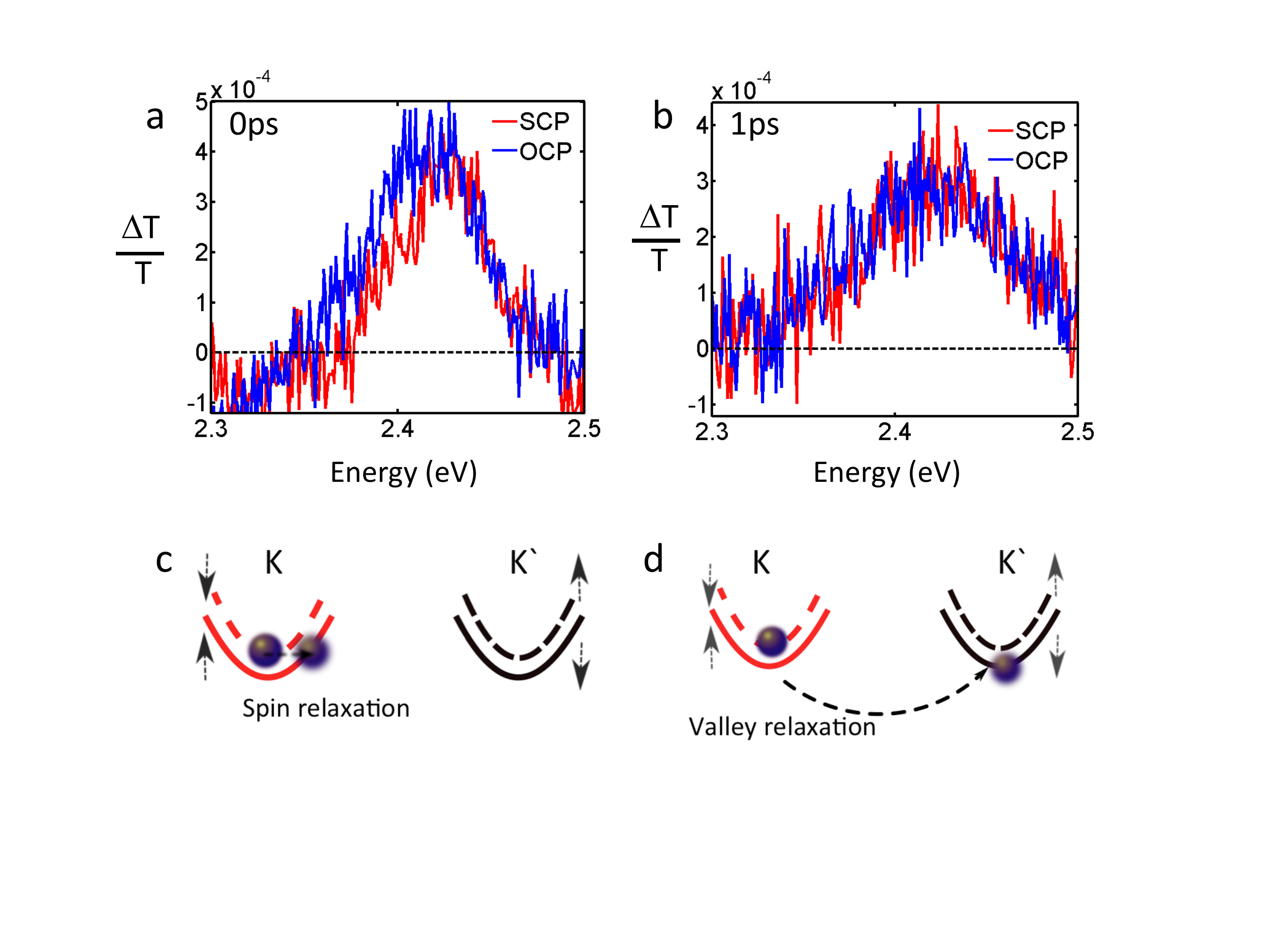}
\caption{(Color online) (a-b) Differential transmission of B exciton peak at
various delay at 110K,  0fs (a) and 1ps (b) delays. Red is SCP and blue is
OCP (c-d) Schematic illustration of spin and valley relaxation of electron.} 
\end{figure}
\end{center}

\clearpage

\begin{center}
\begin{figure}[tbp]
\includegraphics[width=7cm,angle=0]{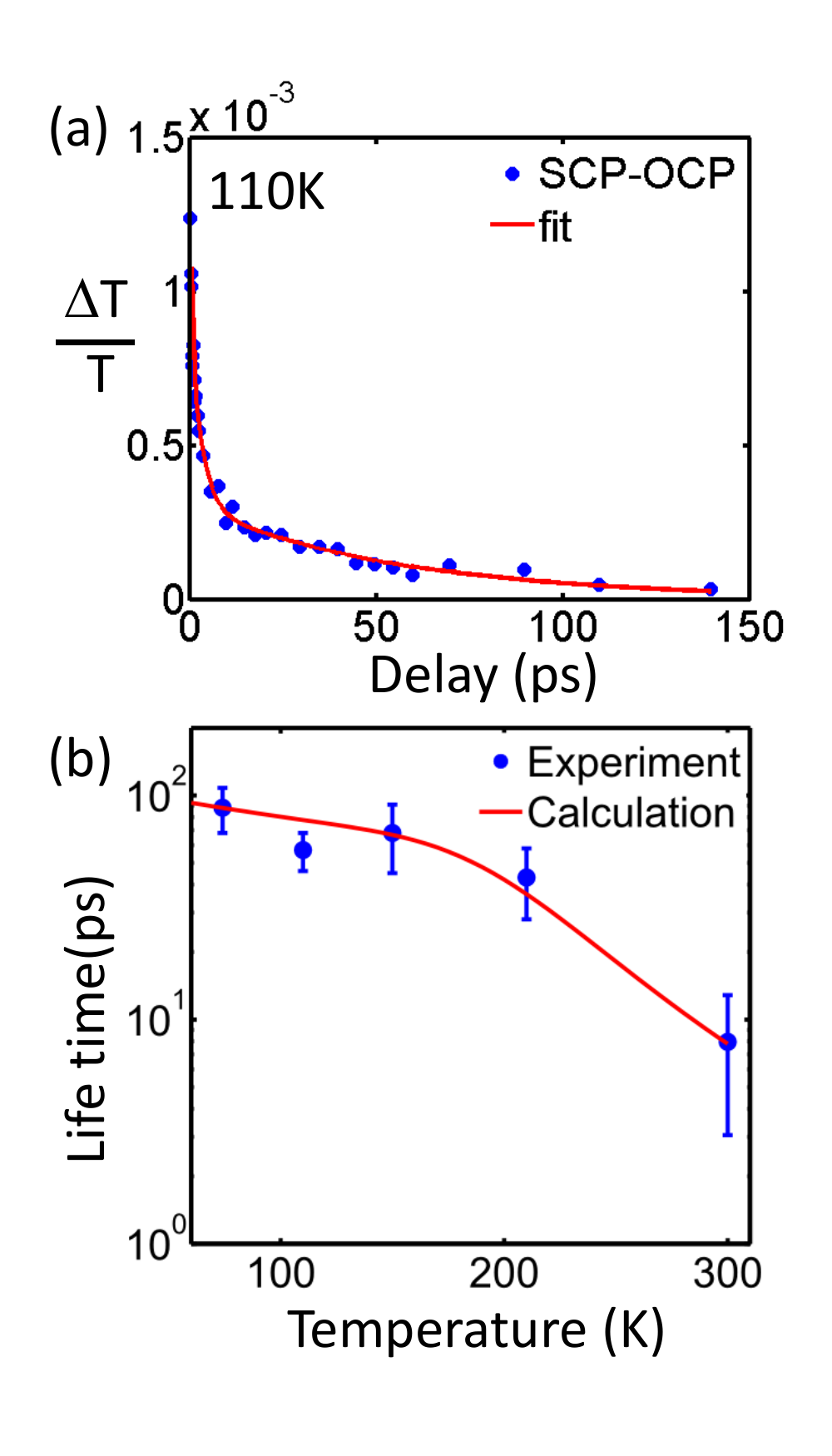}
\caption{(Color online) (a) time evolution of SCP-OCP peak intensity at 110 K and the fit to the three exponential decay function (solid line). (b) The temperature dependence of the slowest time constant.}
\end{figure}
\end{center}

\clearpage


\clearpage

\begin{thebibliography}{99}
\bibitem{Novoselov05} K. S. Novoselov, D. Jiang, F. Schedin, T. J. Booth, V.
V.Khotkevich, S. V. Morozov, and A. K. Geim, Proc. Natl. Acad. Sci. U.S.A.
\textbf{102}, 10451 (2005).

\bibitem{Geim09} A. K. Geim, Science \textbf{324}, 1530 (2009).

\bibitem{Butler13} S. Z. Butler, S. M. Hollen, L. Cao, Y. Cui , J. A. Gupta,
H. R. Guti\'{e}rrez, T. F. Heinz, S. S. Hong , J. Huang, A. F. Ismach, E.
Johnston-Halperin, M. Kuno, V. V. Plashnitsa , R. D. Robinson, R. S. Ruoff,
S. Salahuddin, J. Shan, L. Shi , M. G. Spencer, M. Terrones, W. Windl, and
J. E. Goldberger, ACS Nano \textbf{7}, 2898 (2013).

\bibitem{JariwalaDevices}D. Jariwala, V. K. Sangwan, L. J. Lauhon, T. J. Marks, and M. C. Hersam, ACS Nano \textbf{8}, 1102 (2014).

\bibitem{Wang12} Q. H. Wang, K. Kalantar-Zadeh, A. Kis, J. N. Coleman, and
M. S. Strano, Nat. Nanotechnol. \textbf{7}, 699 (2012).

\bibitem{Radisavljevic11} B. Radisavljevic, A. Radenovic, J. Brivio, J. V.
Giacometti, and A. Kis, Nat. Nanotechnol. \textbf{6}, 147 (2011).

\bibitem{Britnell13} L. Britnell, R. M. Ribeiro, A. Eckmann, R. Jalil, B. D.
Belle, A. Mishchenko, Y.-J. Kim, R. V. Gorbachev, T. Georgiou, S. V.
Morozov, A. N. Grigorenko, A. K. Geim, C. Casiraghi, A. H. Castro Neto, and
K. S. Novoselov, Science \textbf{340}, 1311 (2013).

\bibitem{Sundaram13} R. S. Sundaram , M. Engel, A. Lombardo, R. Krupke, A.
C. Ferrari, Ph. Avouris, and M. Steiner, Nano Lett. \textbf{13}, 1416 (2013).

\bibitem{Perkins13} F. K. Perkins, A. L. Friedman, E. Cobas, P. M. Campbell,
G. G. Jernigan, and B. T. Jonker, Nano Lett. \textbf{13}, 668 (2013).

\bibitem{He13} K. He, C. Poole, K. F. Mak, and J. Shan, Nano Lett. \textbf{13%
}, 2931 (2013).

\bibitem{Splendiani10} A. Splendiani, L. Sun, Y. Zhang, T. Li, J. Kim, C.-Y.
Chim, G. Galli, and F. Wang, Nano Lett. \textbf{10}, 1271 (2010).

\bibitem{Mak10} K. F. Mak, C. Lee, J. Hone, J. Shan, and T. F. Heinz, Phys.
Rev. Lett. \textbf{105}, 136805 (2010).

\bibitem{Xiao12} D. Xiao, G. B. Liu, W. Feng, X. Xu, and W. Yao, Phys. Rev.
Lett. \textbf{108}, 196802 (2012).

\bibitem{YuEWS}Y. Yu, S. Huang, Y. Li , S. N. Steinmann, W. Yang , and L. Cao, Nano Lett., \textbf{14}, 553 (2014).

\bibitem{Mak12} K. F. Mak, K. He, J. Shan, and T. F. Heinz, Nat.
Nanotechnol. \textbf{7}, 494 (2012).

\bibitem{Zeng12} H. Zeng, J. Dai, W. Yao, D. Xiao, and X. Cui, Nat.
Nanotechnol. \textbf{7}, 490 (2012).

\bibitem{Cao12} T. Cao, G. Wang, W. Han, H. Ye, C. Zhu, J. Shi, Q. Niu, P.
Tan, E. Wang, B. Liu and J. Feng, Nat. Commun. \textbf{3}, 887 (2012).

\bibitem{Kosmider13} K. Ko\'{s}mider and J. Fern\'{a}ndez-Rossier, Phys.
Rev. B \textbf{87}, 075451 (2013).

\bibitem{Gutierrez13} H. R. Guti\'{e}rrez, N. Perea-L\'{o}pez, A. L. El\'{\i}%
as, A. Berkdemir, B. Wang, R. Lv, F. L\'{o}pez-Ur\'{\i}as, V. H. Crespi, H.
Terrones, and M. Terrones, Nano Lett. \textbf{13}, 3447 (2013).

\bibitem{Zeng13} H. Zeng, G.-B. Liu, J. Dai, Y. Yan, B. Zhu, R. He, L. Xie,
S. Xu, X. Chen, W. Yao, and X. Cui, Sci. Rep. \textbf{3}, 1608 (2013).

\bibitem{Zhu11} Z. Y. Zhu, Y, C. Cheng and U. Schwingenschlogl, Phys. Rev. B
\textbf{84}, 153402 (2011).

\bibitem{Zhao13} W. Zhao, Z. Ghorannevis, L. Chu, M. Toh, C. Kloc, P.-H.
Tan, and G. Eda, ACS Nano \textbf{7}, 791 (2013).

\bibitem{Jones13} A. M. Jones, H. Yu, N. J. Ghimire, S. Wu, G. Aivazian, J.
S. Ross, B. Zhao, J. Yan, D. G. Mandrus, D. Xiao, W. Yao, and X. Xu, Nat.
Nanotechnol. \textbf{8}, 634 (2013).

\bibitem{Sun13} L. Sun, J. Yan, D. Zhan, L. Liu, H. Hu, H. Li, B. K. Tay,
J.-L. Kuo, C.-C. Huang, D. W. Hewak, P. S. Lee, and Z. X. Shen, Phys. Rev.
Lett. \textbf{111}, 126801 (2013).

\bibitem{Shi13} H. Shi, R. Yan, S. Bertolazzi, G. Gao, A. Kis, D. Jena, H.
Xing and L. Huang, ACS Nano \textbf{7}, 1072 (2013).

\bibitem{Sim13} S. Sim, J. Park, J. Song, C. In, Y. Lee, H. Kim, and H.
Choi, Phys. Rev. B \textbf{88}, 075434 (2013).

\bibitem{Mai14} C. Mai, A. Barrette, Y. Yu, Y. G. Semenov, K. W. Kim, L.
Cao, and K. Gundogdu. Nano Lett., \textbf{14}, 202 (2014).

\bibitem{Wang13} Q. Wang, S. Ge, X. Li, J. Qiu, Y. Ji, J. Feng and D. Sun.
ACS Nano \textbf{7}, 11087 (2013).

\bibitem{YuVapordep}Y. Yu, S. Hu, L. Su, L. Huang, Y. Liu, Z. Jin, A. A. Purezky, D. B. Geohegan, K. W. Kim, Y. Zhang, L. Cao, arXiv:1403.6181 (2014)

\bibitem{S1}
Y. -H. Lee, L. L. Yu. H. Wang, W. J. Fang, X. Ling, Y. M. Shi, C. T. Lin, J. K. Huang, M. T. Chang, C. S. Chang, M. Dresselhaus, T. Palacios, L. J. Li, and J. Kong, Nano Lett. \textbf{13}, 1852 (2013). 

\bibitem{S2}
S. Najmaei, Z. Liu, W. Zhou, X. Zou, G. Shi, S. Lei, B. I. Yakobson, J.-C. Idrobo, and P. M. Ajayan and J. Lou, Nat. Mater. \textbf{12}, 754 (2013).

\bibitem{S3}
A. M. van der Zande, P. Y. Huang, D. A. Chenet, T. C. Berkelbach, Y.M. You, G.-H. Lee, T. F. Heinz, D. R. Reichman, D. A. Muller and J. C. Hone, Nat. Mater. \textbf{12}, 554 (2013).

\bibitem{Sallen12} G. Sallen, L. Bouet, X. Marie, G. Wang, C. R. Zhu, W. P.
Han, Y. Lu, H. Tan, T. Amand, B. L. Liu, and B. Urbaszek, Phys. Rev. B \textbf{86}, 081301, (2012).

\bibitem{Lee86} Y. H. Lee, A. Chavez-Pirson, S. W. Koch, H. M. Gibbs, S. H.
Park, J. Morhange, A. Jeffery, and N. Peyghambarian, Phys. Rev. Lett.
\textbf{57}, 2446 (1986).

\bibitem{Christodoulides10} D. N. Christodoulides, I. C. Khoo, G. J. Salamo,
G. I. Stegeman, and E. W. V. Stryland, Adv. Opt. Photonics \textbf{2}, 60
(2010).

\bibitem{Kuc11} A. Kuc, N. Zibouche, and T. Heine. Phys. Rev. B \textbf{83},
245213 (2011).

\bibitem{Peyghambarian84} N. Peyghambarian, H. M. Gibbs, J. L. Jewell, A.
Antonetti, A. Migus, D. Hulin, and A. Mysyrowicz Phys. Rev. Lett. \textbf{53}%
, 2433 (1984).

\bibitem{Zhu14} B. Zhu, X. Chen, X. Cui, arXiv:1403.5108v1 (2014).

\bibitem{Lu13} H. Lu, W. Yao, D. Xiao, S. Shen, Phys.
Rev. Let. \textbf{110}, 016806, (2013).

\bibitem{S4}
P. Giannozzi et al., J. Phys.: Condens. Matter \textbf{21}, 395502 (2009). 

\bibitem{S5}
A. Klein, S. Tiefenbacher, V. Eyert, C. Pettenkofer, and W. Jaegermann, Phys. Rev. B \textbf{64}, 205416 (2001). 

\bibitem{S6}
H.-P. Komsa, and A.V. Krasheninnikov, Phys. Rev. B \textbf{88}, 085318 (2013).

\bibitem{S7}
X. Li, J., T. Mullen, Z. Jin, K.M. Borysenko, M. Buongiorno Nardelli, and K.W. Kim, Phys. Rev. B \textbf{87}, 115418 (2013)

\bibitem{Song13} Y. Song, and H. Dery. Phys. Rev. Let. \textbf{111}%
,026601, (2013).

\bibitem{Hall03} K. C. Hall, K. G\"{u}ndogdu, E. Altunkaya, W. H. Lau, M. E.
Flatt\'{e}, T. F. Boggess, J. J. Zinck, W. B. Barvosa-Carter, and S. L.
Skeith, Phys. Rev. B \textbf{68}, 115311 (2003).

\bibitem{Zerrouati88} K. Zerrouati, F. Fabre, G. Bacquet, J. Bandet, J.
Frandon, G. Lampel, and D. Paget, Phys. Rev. B \textbf{37}, 1334 (1988).

\bibitem{Dyakonov71} M. I. D'yakonov and V. I. Perel, Zh. Eksp. Teor. Fiz.
60, 1954 (1971) [Sov. Phys. JETP \textbf{38}, 1053].

\bibitem{Yafet52} Y. Yafet, Phys. Rev. \textbf{85}, 478 (1952).

\bibitem{Elliott54} R. J. Elliott, Phys. Rev. \textbf{96}, 266 (1954).
\end{thebibliography}
\end{document}

\author{Cong Mai}
\affiliation{Department of Physics, North Carolina State University, Raleigh, NC 27695, USA}
\title{Supplementary Information: Temperature Dependent Valley Relaxation Dynamics in Single Layer $WS_{2}$ Measured Using Ultrafast Spectroscopy}
\author{Yuriy G. Semenov}
\affiliation{Department of Electrical and Computer Engineering, North Carolina State University, Raleigh, NC 27695, USA}
\author{Andrew Barrette}
\affiliation{Department of Physics, North Carolina State University, Raleigh, NC 27695, USA}
\author{Yifei Yu}
\affiliation{Department of Material Science and Engineering, North Carolina State University, Raleigh, NC 27695, USA}
\author{Zhenghe Jin}
\affiliation{Department of Electrical and Computer Engineering, North Carolina State University, Raleigh, NC 27695, USA}
\author{Linyou Cao}
\affiliation{Department of Material Science and Engineering, North Carolina State University, Raleigh, NC 27695, USA}
\author{Ki Wook Kim}
\affiliation{Department of Electrical and Computer Engineering, North Carolina State University, Raleigh, NC 27695, USA}
\author{Kenan Gundogdu}
\affiliation{Department of Physics, North Carolina State University, Raleigh, NC 27695, USA}

\pacs{72.25.-b, 75.85.+t, 73.50.Jt, 85.75.-d }
%\pacs{71.15.Mb, 72.10.Di, 72.20.Ht, 73.50.Dn}
\maketitle

\section{SAMPLE GROWTH METHOD}
Samples were grown by chemical vapor deposition (CVD) in a tube furnace (Lindberg/Blue M$^\text{TM}$ 1100 $^{\circ}$C Tube Furnaces from Thermo Scientific, Inc) \cite{S1,S2,S3}. An Al$_2$O$_3$ crucible with sulfur powder was loaded at the upstream (outside of the furnace heating zone) and WO$_3$ source material was placed on a Si substrate at the center of the furnace. Several double polished sapphire substrates were placed at the downstream (close to the source materials) in the quartz tube. The temperature was ramped to 550 $^{\circ}$C in 30 min, then gradually increased to 900 $^{\circ}$C in 80 min and kept for 10 min. The whole process was under an atmosphere of forming gas (5\% H$_2$, 95\% Ar) with a flux rate of 100 sccm. After that, it was allowed to cool down to room temperature naturally. 

\section{EXPERIMENTAL DATA OF DECAY IN VALLEY POLARIZATION}
Fig. S1 shows circularly polarized anisotropy decay at transient absorption spectra of A exciton peak at 5 different temperatures (74K,110K, 150K, 210K and 300K. Red solid lines are triple exponential fits to the data. The time constants are shown in Table S1. Fig. S2 shows the temperature dependence of the obtained time constants.

\begin{center}
\begin{figure}
\includegraphics[width=15cm]{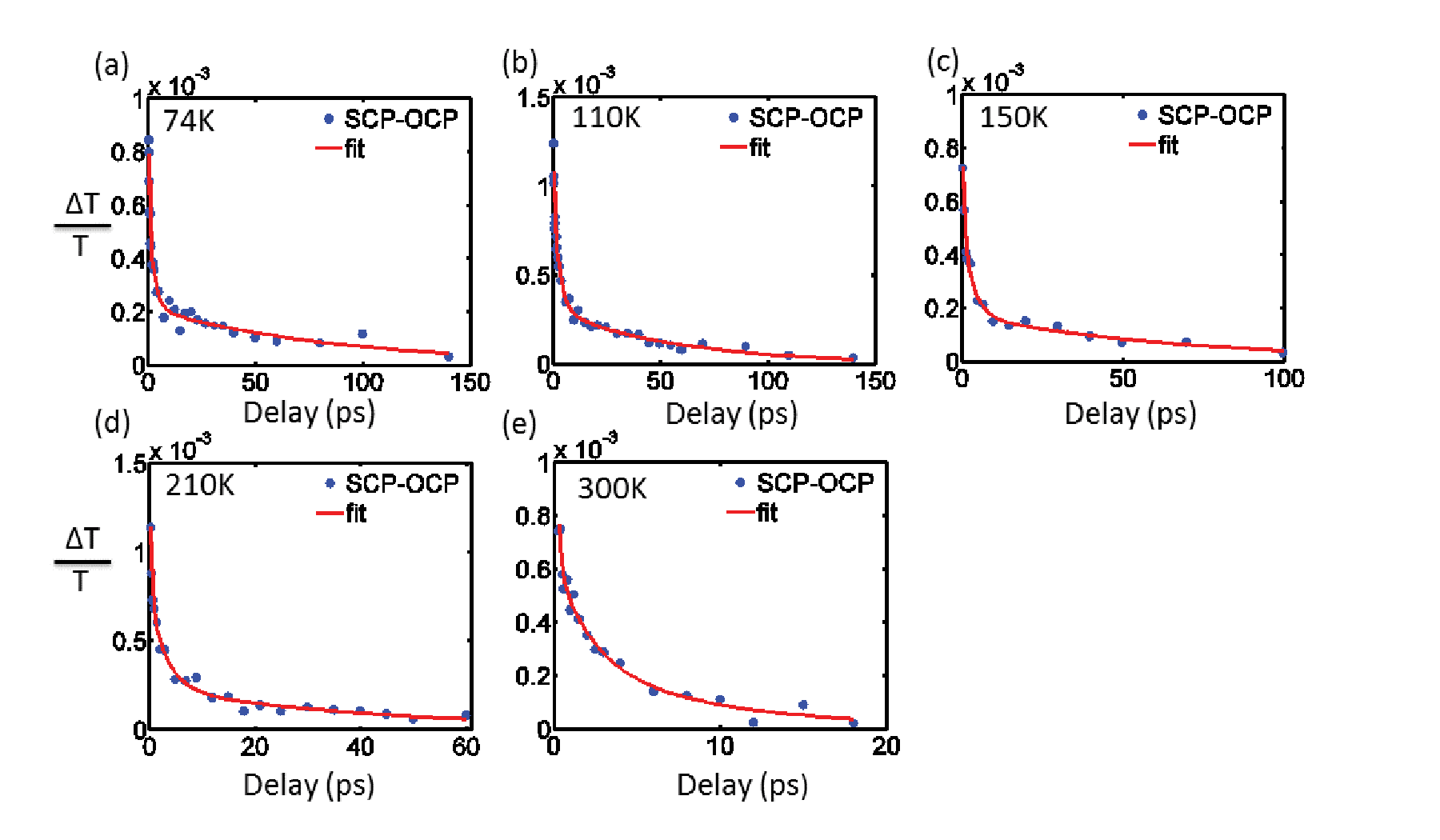} %[bb=16 388 543 684, width=8.5cm]
\caption{(Color online) (a)-(e) SCP-OCP dynamics versus delay at various temperatures.    } \label{Structure}
\end{figure}
\end{center}

\begin{table}[ht]
\setlength{\tabcolsep}{12pt}
\caption{Parameters in triple exponential fitting of SCP-OCP. The model we employed is $A(t)=A_1 e^{(-t/\tau_1 )} +A_2 e^{(-t/\tau_2 )} +A_3 e^{(-t/\tau_3 )}$ . Here $\tau_1$,$\tau_2$,$\tau_3$ refer to three different time constants in decay of valley polarization respectively. We also list the relative contribution of these three different components. The error bar refers to standard error.} % title of Table
\vspace{1cm}
\centering % used for centering table
\begin{tabular}{c c c c c c c} % centered columns (4 columns)
\hline\hline %inserts double horizontal lines
Temp. (K) & $\tau_1$ (ps) & $\tau_2$ (ps) & $\tau_3$ (ps) & $A_1$ (\%) & $A_2$ (\%)& $A_1$ (\%) \\
%heading
\hline % inserts single horizontal line
74  & 0.3$\pm$0.1 & 2.6$\pm$0.8 & 88$\pm$20 & 57 & 27 & 16 \\ % inserting body of the table
110  & 0.2$\pm$0.1 & 3.1$\pm$0.6 & 57$\pm$11 & 54 & 30 & 15 \\ 
150  & 0.3$\pm$0.1 & 3.3$\pm$1.1 & 68$\pm$23 & 54 & 30 & 15 \\ 
210  & 0.2$\pm$0.1 & 3.1$\pm$0.9 & 43$\pm$15 & 55 & 31 & 14 \\ 
300  & 0.1$\pm$0.0 & 1.6$\pm$1.4 & 8$\pm$5 & 83 & 9 & 8 \\ 
\hline %inserts single line
\end{tabular}
\label{table:nonlin} % is used to refer this table in the text
\end{table}

\clearpage
\begin{center}
\begin{figure}
\includegraphics[width=15cm]{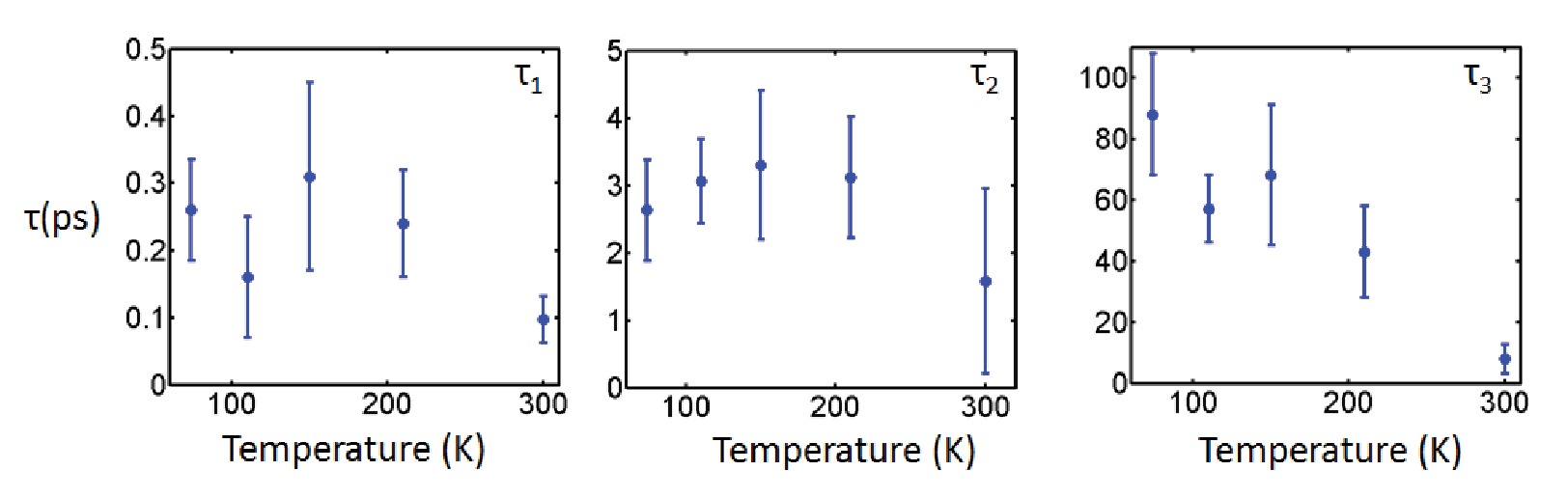} %[bb=16 388 543 684, width=8.5cm]
\caption{(Color online) Temperature dependence of valley lifetime. $\tau_1$,$\tau_2$,$\tau_3$ refer to time constants in triple exponential fitting. In particular, $\tau_1$ (left panel) refers to fastest decay, $\tau_3$ refer to slowest decay and $\tau_2$ refer to intermediate decay. (See Table S1.) } \label{Structure}
\end{figure}
\end{center}

\section{\protect\bigskip Band structure calculation}
To examine the accuracy of band alignments and spin orbic coupling effect, the Density Functional Theory (DFT) calculations were performed with norm conserving pseudopotentials and local-density approximation is used for the exchange-correlation functional and, as it is implemented in the QUANTUM ESPRESSO package~\cite{S4}. The optimum atomic configurations were found through structural relaxation to minimize interatomic forces and total system energy, which were used for following electronic band calculations. For the k-point sampling, 18 $\times$ 18 $\times$ 1 Monkhorst-Pack grid were used with energy cut-off 60 Ry in the plane-wave expansion. 

In our calculation, the spin-orbit coupling is included in the pseudopotential we build, the optimized lattice constant for monolayer WS$_2$ is 3.13 \AA, consistent with other theoretical studies~\cite{S5,S6}. Fig. S3 shows the band structure of monolayer WS$_2$.  The calculated direct band gap whose magnitudes is 1.65 eV, that is a little lower than experiment result. The spin-orbit interaction leads to a splitting of 434 meV at the valence band and 34 meV at in the valence and conduction bands, respectively, at the  of K- point.

\begin{center}
\begin{figure}
\includegraphics[width=8cm]{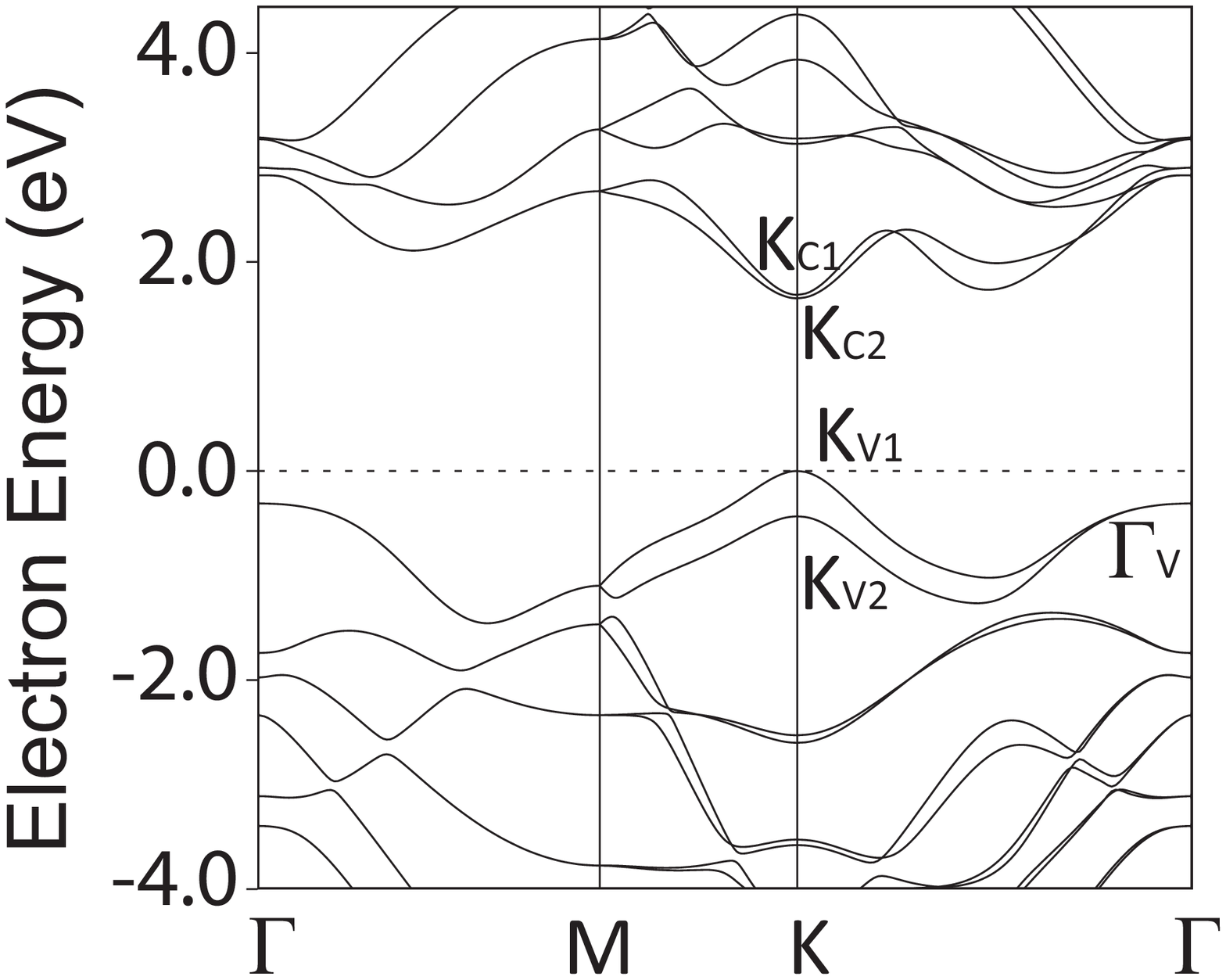} %[bb=16 388 543 684, width=8.5cm]
\caption{Electronic band structures of WS$_2$ monolayer along the symmetry directions in the FBZ, with spin-orbit coupling effect. The energy scale is adjusted using valence band maximum at the point serves as the reference. The related energy level at the symmetry points are K$_{\text{C}1}$: 1.686eV, K$_{\text{C}2}$: 1.652eV, K$_{\text{V}1}$: 0.000eV, K$_{\text{V}2}$: -0.434eV, $\Gamma_{\text{V}}$: -0.310eV. } \label{Structure}
\end{figure}
\end{center}

\section{\protect\bigskip Rashba effect from first principal calculation }

Using first principal calculation, we can provide the evidence of Rashba effect by calculating the in-plane spin splitting around K and K' region under vertical electrical field. 
Beyond the first principle calculation of band structure, vertical saw-like electrical field is applied to the system for bringing the chemical potential difference between two S atoms. The system is relaxed and the spin expectation value $\langle  S_x \rangle$,$\langle  S_y \rangle$ are calculated around K and K' points under situations with and without electric field separately. The spin expectation value are zero in the calculated region without electric field. After applying the electric field, $\sqrt{ \langle  S_x \rangle ^2+\langle  S_y \rangle ^2}$ value remains zero only at the K and K' points and it gets increased linearly along the K-M line, as shown in Fig. S4. 

\begin{center}
\begin{figure}
\includegraphics[width=8cm]{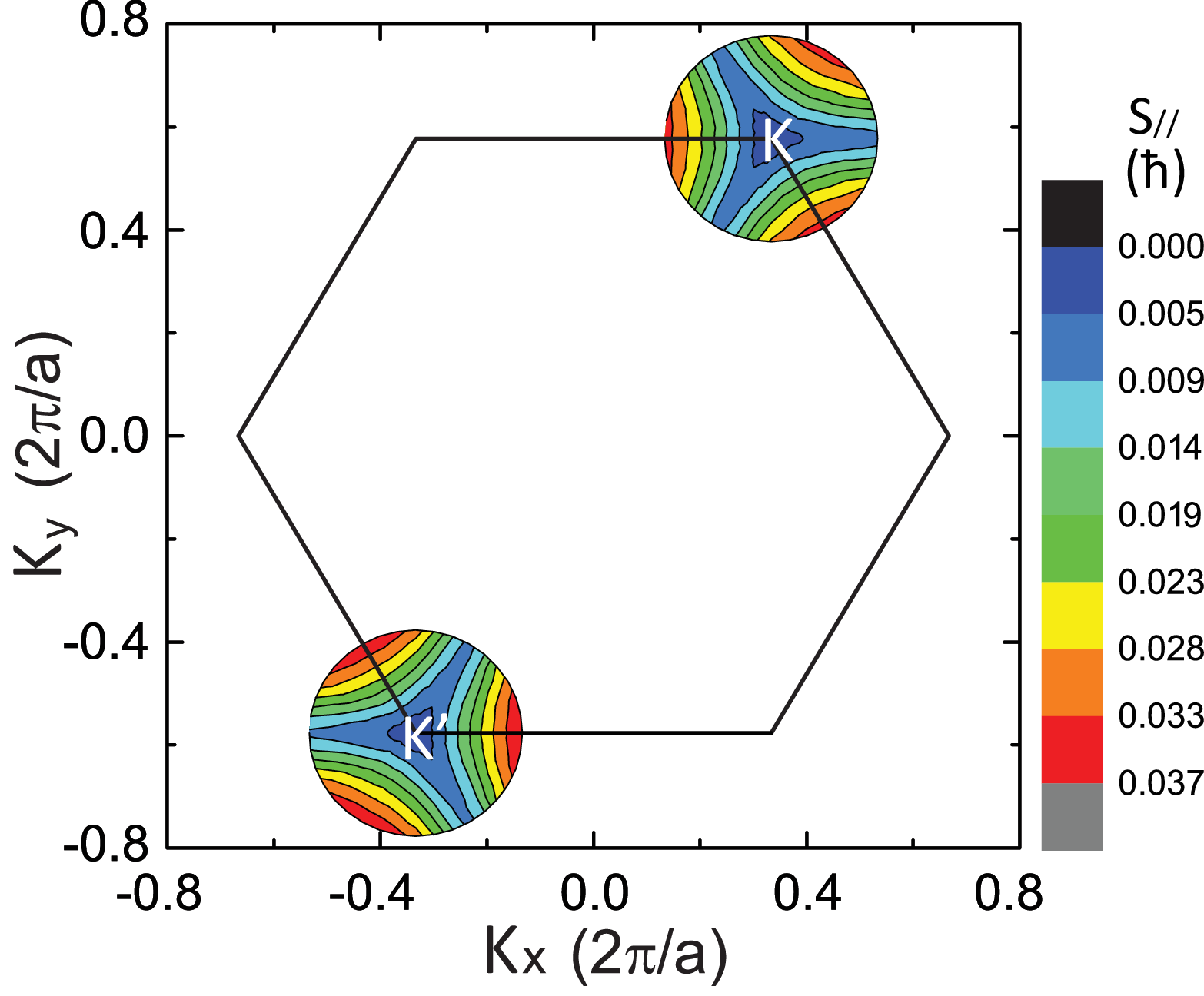} %[bb=16 388 543 684, width=8.5cm]
\caption{(Color online) Expectation value of in-plane spin under electric field 10$^{10}$ V/m. The results are in units of $\hbar$. The chemical potential difference between S atoms is 3.08eV. } \label{Structure}
\end{figure}
\end{center}

\section{phonon-assistant momentum relaxation rate from first principal calculation }

Electron-phonon scattering process can be described and simulated by first principal calculations as well~\cite{S7}. The calculation are based on electronic and phononic band structure of monolayer WS$_2$. The intra-band scattering rate can be described by using Fermi's golden rule. The scattering rates are affected by temperature as well as the hole energy. To simplify the calculation, we pick the data point where the hole energies are near zero. The scattering rates as a function of temperature as listed in Table S2. The data show great linearity within temperature range from 70K to 200K. 
\begin{table}[ht]
\setlength{\tabcolsep}{12pt}
\caption{Scattering rates of monolayer WS$_2$ due to the hole-phonon interactions from first principle calculation} % title of Table
\vspace{1cm}
\centering % used for centering table
\begin{tabular}{c c} % centered columns (4 columns)
\hline\hline %inserts double horizontal lines
Temperature & Scattering Rate\\ [-6pt]
(K) & ($s^{-1}$)  \\% inserts table
%heading
\hline % inserts single horizontal line
70  & 1.03$\times 10^{13}$  \\ % inserting body of the table
100 & 1.11$\times 10^{13}$  \\
130 & 1.22$\times 10^{13}$ \\
170& 1.39$\times 10^{13}$  \\
200& 1.54$\times 10^{13}$  \\ [1ex] % [1ex] adds vertical space
\hline %inserts single line
\end{tabular}
\label{table:nonlin} % is used to refer this table in the text
\end{table}